\documentstyle[twocolumn,aps,graphicx,epsfig]{revtex}

\begin{document}

\wideabs{\title{Biased EPR entanglement and its application
to teleportation}

\author{W.~P.~Bowen, P.~K.~Lam}

\address{Department of Physics, Faculty of Science, \\
The Australian National University, A.C.T. 0200, Australia.}

\author{T.~C.~Ralph}

\address{Department of Physics, Centre for Quantum Computer Technology,\\
University of Queensland, St Lucia, QLD, 4072 Australia}
\date{\today}
\maketitle

\begin{abstract}
We consider pure continuous variable entanglement with non-equal
correlations between orthogonal quadratures.  We introduce a simple
protocol which equates these correlations and in the process
transforms the entanglement onto a state with the minimum allowed
number of photons.  As an example we show that our protocol
transforms, through unitary local operations, a single squeezed beam
split on a beam splitter into the same entanglement that is produced
when two squeezed beams are mixed orthogonally.  We demonstrate that
this technique can in principle facilitate perfect teleportation
utilising only one squeezed beam.
\end{abstract}}
\narrowtext

\section{Introduction}
Entanglement is a key ingredient in many quantum communication tools
\cite{Bennett4}.  It has been studied in dichotomic regimes such as
photon-counting \cite{Discrete} and electron spin \cite{Spin} and also
in continuous variable regimes \cite{Einstein,Reid}.  Recently many of
the quantum information protocols and tools developed dichotomically
have been generalised to the continuous variable regime
\cite{Bra98,Lloyd,Polk,ral99,Duan} and new experimental measures of
continuous variable entanglement have also been proposed
\cite{Bana,Ral00,Duan2}.

The most studied and generated form of entanglement in continuous
variable quantum optics is Einstein-Podolsky-Rosen (EPR) entanglement
\cite{Reid,Kim}.  EPR entanglement is characterised by quantum
correlations between conjugate quadrature amplitudes of two light
beams.  For example there may be quantum correlations between both the
amplitude and the phase quadratures of the two beams.  Usually EPR
entanglement in which the correlations between conjugate quadratures
are of equal strength is discussed.  We will refer to this as {\it
unbiased} EPR entanglement.  There is no reason to restrict ourselves
to this case however, and recently van Loock and Braunstein
\cite{Loock} examined entanglement with biased correlations.  They
evaluated the entanglement in terms of continuous variable
teleportation of coherent states, studying two party and multi-party
protocols.

In this paper we examine this type of entanglement in a more general
way.  Using a standard measure of EPR entanglement and introducing a
new measure based on the number of photons present in a state, we show
in what sense unbiased EPR entanglement is maximal.  A key issue in
our treatment is that of resources.  That is, the number of photons
used to produce a particular level of entanglement.  Although an
integral part of discrete variable discussions of quantum information,
the question of resources is not always explicit in continuous
variable treatments.  Here we define maximal EPR states as those that
use the minimum resources (photons) necessary to produce a particular
degree of correlation.

We then discuss how it is possible to move between biased and unbiased
states using only local operations.  Using the example of continuous
variable teleportation we show that an improvement in efficacy occurs
with our protocol.  In general unbiased entanglement provides optimum
results, however we also show that there are some special situations
where biased entanglement is optimal.

Protocols for manipulating entanglement are common in discrete 
variable quantum information.  Perhaps most useful is distillation, or 
as it is sometimes called, concentration of entanglement 
\cite{Bennett2,Bennett3}.  Distillation is a process involving only 
local operations and classical communication which takes some number 
of weakly entangled particles to a smaller number of more entangled 
particles.  The total entanglement cannot be increased by such a 
process, only the entanglement per particle.  Continuous variable 
distillation protocols \cite{Duan,Parker} take some number of weakly 
entangled modes and produce a smaller number of more entangled modes.  
In contrast the operations desribed in this paper involve single mode 
manipulations.

The paper is arranged in the following way: Section II reviews the
techniques for characterising and producing EPR entanglement and
introduces EPR symmetrisation and the concept of non-maximal EPR
states for continuous variables.  In Section III our protocol to
redistribute the quantum correlations is introduced and in Section IV
we consider its application to teleportation as an example.  The
protocol is generalised to tri-partite entanglement in Section V and
we conclude in Section VI.

\section{Continuous variable entanglement}

\subsection{Characterisation}
An optical beam can be expressed as a mean unchanging field plus a
fluctuation term \cite{Walls}.
\begin{equation}
	 \hat{A}(t) = \langle \hat{A} \rangle + \delta \!  \hat{A}(t) =
	 \langle \hat{A} \rangle + \frac{1}{2}(\delta \!
	 \hat{X}(t)^{+} + i \ \delta \!  \hat{X}(t)^{-})
	 \label{generalbeam}\\
\end{equation}
Here $\hat{A}(t)$ is a field annihilation operator and $\langle \hat A
\rangle$ is its expectation value.  $\delta \!  \hat{A}(t)$ contains
all of the fluctuations of the operator and can be split into two
orthogonal hermitian operators, quadrature phase $\delta \!
\hat{X}(t)^{-}$ and quadrature amplitude $\delta \!  \hat{X}(t)^{+}$.
The superscripts $-$ and $+$ distinguish the phase and amplitude
quadratures, respectively.

Taking the Fourier transform of $\hat{A}(t)$ produces the frequency
domain operator $A(\Omega)$ which we denote by removing the
hat.  The variance of the phase and amplitude quadratures of an
optical beam are given by $V^{\pm} = V^{\pm}(\Omega) = V (\delta \!
X^{\pm}(\Omega) ) = \langle (\delta \!  {X}^{\pm}(\Omega))^{2} \rangle
$.  Throughout this paper we will consider sidebands at frequencies
$\pm \Omega$ away from the carrier frequency and henceforth the
frequency $\Omega$ will not be shown explicitly.  The Heisenberg
uncertainty relationship can then be expressed in terms of these
variances as $V^{+} V^{-} \ge 1$.

In this paper the degree of quantum correlation between a given
entangled pair of beams is measured with the product of conditional
variances, $V_{\rm cv}^{\pm}$, of conjugate quadratures between the
two beams.  This measure was first proposed by Reid and Drummond
\cite{Reid}, is easily measureable, and has been used to characterise
entanglement in a number of experiments \cite{Kim,SHANGXI,Silber}.
The conditional variance measures how well a quadrature amplitude of
one beam can be inferred from a quadrature measurement of the other
beam, or in other words it measures the variance of the noise
degrading the otherwise perfect correlations between the beams.  By
determining the conditional variances of two conjugate quadratures
between beams, we may demonstrate the EPR-paradox experimentally
\cite{Kim}.  This demonstration is a sufficient but not necessary
condition for the states to be entangled (i.e inseparable).  Limiting
ourselves to the amplitude and phase
quadratures, the condition for EPR entanglement becomes
\begin{equation}
	\label{eqEPR4}
	V_{\rm cv}^{+}V_{\rm cv}^{-} < 1
\end{equation}
where the conditional variances are given by $ V_{\rm cv}^{\pm} =
V^{\pm}_{\rm b} - | \langle \delta \!{X}^{\pm}_{\rm b} \delta \!
{X}^{\pm}_{\rm a} \rangle |^{2}/V^{\pm}_{\rm a} $ \cite{holl} and the
subscripts a and b label the two optical beams.  $V_{\rm cv}^{+}V_{\rm
cv}^{-} = 1$ defines a hard boundary below which the state must be
entangled, the closer $V_{\rm cv}^{+}V_{\rm cv}^{-}$ is to $0$ the
stronger the entanglement.
%
%
It is also possible to have rotated EPR correlations such
that the conditional variance between (say) the amplitude of one beam
and the phase of the other, and vice versa obey relations analogous to
Eq.(\ref{eqEPR4}).  We will only consider non-rotated states here.

%
A number of measures of entanglement are in useage.  In particular the 
Duan criteria \cite{Duan2,Cirac} gives neccessary and sufficient 
conditions for separability of Gaussian states.  We use the EPR 
condition here because of its physical significance and links with the 
efficacy of continuous variable quantum information protocols 
\cite{grang}.  We mainly restrict ourselves in this paper to pure 
states and exclusively to Gaussian states.  For pure Gaussian states 
the EPR condition is qualitatively equivalent to Duan's criteria.

\subsection{Production}
\begin{figure}[ht]
  \begin{center}
  \includegraphics[width=7.5cm]{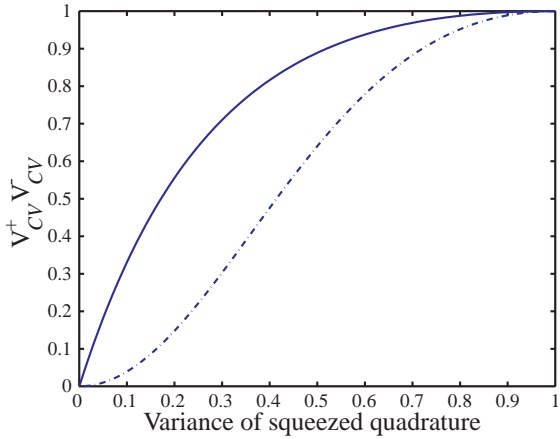}
  \end{center}
  \caption{Entanglement produced with varying degrees of squeezing,
  the solid line represents entanglement from two squeezed beams and
  the dot-dashed line entanglement from only one beam.}
  \label{singledoubleSqueezedEPR}
\end{figure}

Continuous variable entanglement of the EPR type may be produced by
mixing two squeezed beams on a 50/50 beam splitter
\cite{yeo93,Ralph,Furu}. Squeezed beams are ones for which
$V^{+}<1<V^{-}$ or vice versa \cite{Walls}.
Throughout this paper we denote the input
beams to this beam splitter by the subscripts 1 and 2, and the output
beams by EPR1 and EPR2.  For zero phase difference between the inputs
at the beam splitter, the output quadrature amplitude relations are
given by
\begin{eqnarray}
	\label{beamsplitter1}
\delta \!  {X}_{\rm EPR1}^{\pm} & = & \frac{1}{\sqrt{2}}
(\delta \!  {X}^{\pm}_{1} + \delta \!  {X}^{\pm}_{2}) \\
\delta \!  {X}_{\rm EPR2}^{\pm} & = & \frac{1}{\sqrt{2}}
(\delta \!  {X}^{\pm}_{1} - \delta \!  {X}^{\pm}_{2})
\label{beamsplitter2}
\end{eqnarray}
and the conditional variances between these two outputs are
\begin{equation}
	V^{\pm}_{\rm cv} = \frac{2V^{\pm}_{1}V^{\pm}_{2}}{V^{\pm}_{1} +
V^{\pm}_{2}}
\end{equation}
If both input beams (1 and 2) are pure (minimum uncertainty) states,
then
\begin{equation}
\label{cv}
V_{\rm cv}^{+}V_{\rm
cv}^{-}=\frac{4}{2+V^{+}_{1}V^{-}_{2}+V^{-}_{1}V^{+}_{2}}
\end{equation}
Now if both input beams are equally squeezed and the squeezing is in 
orthogonal directions (eg $V^{+}_{1}=V^{-}_{2}<1$ or 
$V^{-}_{1}=V^{+}_{2}<1$) then the output beams will be in the usual, 
unbiased, EPR entangled state.  If the squeezing is not equal for the 
two beams and/or is not on orthogonal quadratures then biased 
entanglement will be produced.  An interesting case is when only one 
input is squeezed (say $V^{+}_{1}<1<V^{-}_{1}$ and $V^{\pm}_{2}=1$).  
From equation (\ref{cv}) we find that $V_{\rm cv}^{+}V_{\rm cv}^{-}$ 
is still less than 1 and thus the outputs are still entangled in this 
case.  In this paper we will use this entanglement as our example, 
however the analysis and techniques we discuss can equally be applied 
to general biased entanglement.  Clearly, for a fixed amount of 
squeezing, two squeezed beams will produce stronger correlations than 
one.  Figure \ref{singledoubleSqueezedEPR} shows $V_{\rm cv}^{+}V_{\rm 
cv}^{-}$ for variable squeezing in both cases.  As the squeezing 
strength increases, the biased and unbiased entanglement both approach 
perfect, $V_{\rm cv}^{+}V_{\rm cv}^{-} \rightarrow 0$.
As well as access to stronger correlations, entanglement from two
squeezed beams has another advantage.  Many quantum information
protocols require quantum correlations in two conjugate quadratures,
$V_{\rm cv}^{+}, V_{\rm cv}^{-} < 1$, which this entanglement
naturally provides.  Entanglement from only one squeezed beam on the
other hand provides quantum correlations in only one quadrature, say
$V_{\rm cv}^{+}<1$ but necessarily $V_{\rm cv}^{-} >1$.  One such
example, is quantum teleportation of a coherent state where perfect
entanglement from one squeezed beam ($V_{\rm cv}^{+}V_{\rm cv}^{-}
=0$) provides a fidelity of only $F = 1/\sqrt{2}$ compared with $F=1$
for perfect unbiased entanglement \cite{Loock}.  In section III we
will show how perfect teleportation can be achieved utilising only one
squeezed beam but first we would like to study the difference between
biased and unbiased entanglement in more detail.

\subsection{EPR Symmetrisation}
We now use a resource argument to define and quantify the difference
between a maximal EPR state and a non-maximal EPR. We will identify
unbiased entanglement as maximal.

First let us consider what we mean by ``maximal'' EPR. Clearly the
{\it maximum} EPR state is the one for which
$V_{cv}^{+}=V_{cv}^{-}=0$.  But creating such a state would require
infinite energy, an unphysical limit.  Thus the idea of a maximal EPR
state should be considered within some limit on resources.  Let us
consider the minimum number of photons required to achieve a
particular value of $V_{\rm cv}^{+}V_{\rm cv}^{-}$.  We will define a
maximal EPR state as one which achieves this particular value of
entanglement with the minimum photon number.  The average photon number in the
sidebands of an optical beam, taken over some small range of
frequencies for which the quadrature variances are constant, can be
related to these variances (in suitably normalised units) via
\begin{eqnarray}
    \label{n}
n & = & \langle \delta \hat A^{\dagger}(\Omega) \delta \hat A(\Omega)+
\delta \hat A^{\dagger}(-\Omega) \delta \hat A(-\Omega) \rangle \nonumber\\
& = & {{1}\over{2}}(V^{+}(\Omega)+V^{-}(\Omega))-1
\end{eqnarray}
The quadrature variances of the individual EPR beams can be related to
the conditional variances between them by invoking the uncertainty
relations.  If it is possible to infer the value of $\delta X_{\rm
EPR1,2}^{+}$ with a variance $V_{\rm cv}^{+}$ then the uncertainty
relation requires that $V_{\rm cv}^{+} V_{\rm EPR1,2}^{-}\ge 1$.
Similarly if we can infer the value of $\delta X_{\rm EPR1,2}^{-}$
with a variance $V_{\rm cv}^{-}$ then the uncertainty relation
requires that $V_{\rm cv}^{-} V_{\rm EPR1,2}^{+}\ge 1$.  Thus
\begin{equation}
    \label{min1}
V_{\rm EPR1,2}^{\pm} \ge {{1}\over{V_{\rm cv}^{\mp}}}
\end{equation}
For pure-state entanglement expressions (\ref{min1}) are true in the
equality, substituting these equalities into equation (\ref{n}) we
obtain the average photon number of each of the EPR beams
\begin{equation}
    \label{nepr}
n_{\rm EPR1,2}={{1}\over{2}}({{1}\over{V_{\rm
cv}^{+}}}+{{1}\over{V_{\rm cv}^{-}}})-1
\end{equation}
For a given degree of quantum correlation ($V_{\rm cv}^{+}V_{\rm
cv}^{-}$) the minimum value of $n_{\rm EPR1,2}$ is obtained when
\begin{equation}
    \label{min2}
    V_{\rm cv}^{+}=V_{\rm cv}^{-}
\end{equation}
and thus when the entanglement is unbiased.  Both equation
(\ref{min2}) and the equality of equation (\ref{min1}) are satisfied
by entanglement produced from two minimum uncertainty equally squeezed
beams.  This identifies such states as maximal EPR states.  Because
for entanglement from a single squeezed beam $V_{\rm cv}^{+}\ne V_{\rm
cv}^{-}$ we can immediately identify it as a non-maximal EPR state.

We describe the maximallity of the EPR entanglement by $\lambda$,
the ratio of the minimum number of photons necessary (for that
entanglement strength) to the number of photons present,
\begin{equation}
    \label{lam}
\lambda={{n_{\rm maximal}}\over{n_{\rm EPR}}}
\end{equation}
where $n_{\rm EPR}$ is the mean photon number of the EPR state being 
analysed and $n_{\rm maximal}$ is the mean photon number of the 
maximal EPR state with the same conditional variance product.  If 
$\lambda=1$ the state is maximal, if $\lambda<1$ the state is 
non-maximal.

The $\lambda$ parameter also identifies mixed states as non-maximal.  
Such states occur when the states used to produce the entanglement are 
not minimum uncertainty, or if the EPR beams suffer loss.  For mixed 
states it is the equality of equation (\ref{min1}) which will not be 
satisfied, whilst the noise degrading the correlations may still be 
unbiased (i.e. $V_{\rm cv}^{+}=V_{\rm cv}^{-}$) .  In the following 
discussion we will not consider mixed states further.

\section{Redistributing the quantum correlations}

We introduce a method to transform pure non-maximal EPR states into
maximal ones.  In particular we will see how biased entanglement can
be transformed into unbiased entanglement using only local unitary
operations.  These operations keep the overall degree of quantum
correlation (as measured by $V_{\rm cv}^{+}V_{\rm cv}^{-}$) constant
whilst increasing $\lambda$ (equation(\ref{lam})).
%
%
\begin{figure}[ht]
  \begin{center}
  \includegraphics[width=6.5cm]{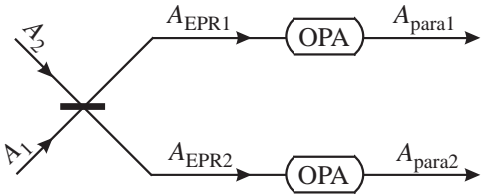}
  \end{center}
  \caption{Scheme to redistribute the quantum noise of an entangled
  pair.}
  \label{OPAdiagram}
\end{figure}

Let us consider the effect of passing entangled beams produced from
only one squeezed beam through two independent degenerate optical
parametric amplifiers (OPAs) as shown in figure \ref{OPAdiagram}. In
the classical pump limit OPAs effect a unitary transformation of the
field \cite{Walls}. The
effect of each OPA is to amplify the fluctuations of one of the input
quadratures whilst de-amplifying the fluctuations of the orthogonal
quadrature.  The quadratures of amplification and de-amplification are
controlled by the phase difference between the input and the pump
field of the OPA. In our example, the amplitude quadratures of both
beams are amplified.  The effect of each OPA on its respective
entangled beam is given by
\begin{eqnarray}
	\label{OPA1}
	\delta \!  {X}^{+}_{\rm para1,2} & = & \sqrt{G} \ \delta \!
	{X}^{+}_{\rm EPR1,2} \\
	\label{OPA2}
	\delta \!  {X}^{-}_{\rm para1,2} & = & \frac{1}{\sqrt{G}} \
	\delta \!  {X}^{-}_{\rm EPR1,2}
\end{eqnarray}
where the gain G has been chosen identically for both OPAs and the
subscripts para1,2 label each entangled beam after the optical
parametric operations.  As the gain terms from each quadrature cancel
on multiplication, these operations leave $V_{\rm cv}^{+}V_{\rm
cv}^{-}$ unchanged.  After the OPAs the entangled quadratures given in
equations (\ref{beamsplitter1}) and (\ref{beamsplitter2}) become
\begin{eqnarray}
	\label{EPRafterOPO1}
	\delta \!  {X}^{+}_{\rm para1} & = & \sqrt{\frac{G}{2}}
	( \delta \!  {X}^{+}_{1} + \delta \!
	{X}^{+}_{2}) \\
	\delta \!  {X}^{-}_{\rm para1} & = & \frac{1}{\sqrt{2G}}
	( \delta \!  {X}^{-}_{1} + \delta \!
	{X}^{-}_{2}) \\
	\label{EPRafterOPO2}
	\delta \!  {X}^{+}_{\rm para2} & =& \sqrt{\frac{G}{2}}
 ( \delta \!  {X}^{+}_{1} - \delta \!
	{X}^{+}_{2}) \\
\delta \!  {X}^{-}_{\rm para2} & =& \frac{1}{\sqrt{2G}} ( \delta \!
{X}^{-}_{1} - \delta \!  {X}^{-}_{2})
\end{eqnarray}

We have previously stated that entanglement generated from one 
squeezed beam has quantum correlations in only one quadrature; two 
squeezed beams generate entanglement that has equal but weaker (for 
the same $V_{\rm cv}^{+}V_{\rm cv}^{-}$) quantum correlations in both.  
For pure entanglement from one squeezed beam setting a gain 
$G=\sqrt{V^{-}_{1}}$ (i.e. equal to the standard deviation of the 
anti-squeezed quadrature of the squeezed input beam) spreads the 
quantum correlations equally over both entangled quadratures.  This 
makes the entanglement unbiased and in fact indistinguishable from 
entanglement generated from two squeezed beams.  More generally a gain 
of
\begin{equation}
	\label{gainchoice}
	G=\sqrt{\frac{V^{-}_{\rm 1}}{V^{+}_{\rm 2}}}
\end{equation}
will make any pure EPR entanglement unbiased and thus maximal.

The OPAs are in fact acting as deamplifiers of photon number in this 
situation.  By choosing the optimum gains we are able to reduce the 
photon number in the beams to the minimum allowed for their particular 
level of entanglement and thereby produce maximal entanglement.  A 
physical picture of this process can be obtained by considering weak 
squeezed beams.  Such beams can be described approximately in the 
Schrodinger picture as superpositions of vacuum and photon pairs: 
$|\psi_{s} \rangle = |0 \rangle + \xi |2 \rangle $ with $\xi<<1$.  
Mixing two of these beams on a beamsplitter with the appropriate phase 
relationship spatially separates the pairs into the individual beams 
producing the maximal entangled state $|\psi_{e} \rangle = |00 \rangle 
+ \xi |11 \rangle $.  On the other hand splitting a single squeezed 
beam gives the state $|\psi_{es} \rangle = |00 \rangle + \xi/\sqrt{2}( 
|11 \rangle +1/2(|20 \rangle+|02 \rangle)) $.  We see that as well as 
the maximal component there is also a component of unseparated pairs.  
These do not contribute to the correlations.  The effect of correctly 
tuned OPAs is to act as 2-photon absorbers which locally remove the 
2-photon components leaving only the maximal component.  A 
generalization of this argument to higher photon number orders can be 
made.

This process is similar to discrete variable concentration protocols 
\cite{Bennett2} since the entanglement per photon is increased.  
Indeed there are strong similarities to the procrustean method of 
discrete concentration, where unentangled photons are ``filtered out'' 
whilst retaining the entangled ones \cite{thew}.  However in the 
discrete case the photon is the fundamental carrier of the 
entanglement whilst in the continuous case it is the field mode which 
plays this role.  In recognition of this distinction we refrain from 
using concentration to describe the continuous variable case.

\section{Teleportation}
In general the goal of quantum teleportation is to destroy a quantum 
state in one system and re-create it exactly in another.  Many 
theoretical papers have been published on quantum teleportation 
\cite{Plenio} and it has been experimentally realised 
\cite{Furu,Bouwm,Boschi}.  The Heisenberg uncertainty principle limits 
the information attainable from any two observables with a non-zero 
commutator relation \cite{Fuchs}.  Thus the perfect reproduction of an 
unknown input state via a direct measurment and recreation scheme is 
impossible.  However a loophole exists if the sender, Alice, and 
receiver, Bob, share entanglement.  This allows them to, in principle, 
teleport the state perfectly.  The uncertainty principle is not 
violated as neither Alice nor Bob learn the identity of the teleported 
state.

\begin{figure}[ht]
  \begin{center}
  \includegraphics[width=7.5cm]{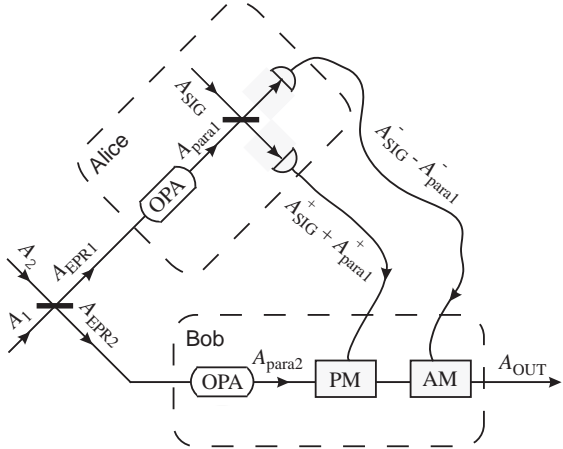}
  \end{center}
  \caption{Quantum teleportation with local OPAs.}
  \label{Telediagram}
\end{figure}

Figure \ref{Telediagram} shows a typical
continuous variable teleportation experiment with the addition of two
local OPAs, one at Alices detection station and one at Bobs
reconstruction station.  The OPAs can conveniently be thought of as a
local resource forming part of the teleportation protocol.  The
subscripts OUT and SIG label the the output and signal beams
respectively.  AM and PM label the amplitude and phase modulators
respectively used to reconstruct the signal state with unity gain on
one of the entangled beams.  The input for the amplitude (phase)
modulator is from an amplitude (phase) detector with suitably chosen
gain.

The overlap integral of the original and teleported state Wigner
functions is a measure of how similar the states are and is
conventionally termed fidelity \cite{Schu}.  Other measures of
teleportation quantify the disturbance (noise) introduced by the
teleporter to an arbitrary teleported state under various conditions
\cite{Ralph}.  For ease of comparison with Ref.\cite{Loock} we will
use fidelity here and only consider unity gain (i.e. the teleportation
gain has been chosen such that the size of the coherent displacement
of the state before and after teleportation are equal).

If all noise sources are Gaussian, the fidelity of coherent state
teleportation at unity gain is given by
\begin{equation}
	F=\frac{2}{\sqrt{ \left (V^{+}_{OUT} + 1 \right ) \left
	(V^{-}_{OUT} + 1 \right )}}
\end{equation}
The fidelity equals 1 when the Wigner function of the output state is
a perfect replica of that of the signal state.  The phase and
amplitude variance of the output state of the teleporter shown in
figure \ref{Telediagram}, assuming it functions perfectly, can be
written as
\begin{eqnarray}
	V^{+}_{OUT} = 2G V^{+}_{1} + V^{+}_{SIG}\\
	V^{-}_{OUT} = \frac{2}{G} V^{-}_{2} + V^{-}_{SIG}
\end{eqnarray}
We have assumed that the signal is coherent thus $V^{+}_{SIG} = V^{-}
_{SIG} = 1$.  This gives a fidelity of
\begin{equation}
	F = \frac{1}{\sqrt{V^{+}_{1} V^{-}_{2} + G V^{+}_{1} +
	\frac{\displaystyle 1}{\displaystyle G} V^{-}_{2} + 1}} \\
	\label{fidelityOPA}
\end{equation}
The maximum fidelity $F_{\rm max}$ for a given entanglement strength
is obtained when the parametric gain $G$ is chosen as in equation
(\ref{gainchoice}) and is given by
\begin{equation}
	F_{\rm max} = \frac{1}{\sqrt{V_{\rm 1}^{+}V_{\rm 2}^{-}}+1}
	\label{Fmax}
\end{equation}
This upper limit of achievable fidelity is set solely by the strength
of the entanglement and, with appropriate parametric gain (dictated by
equation (\ref{gainchoice})), all entanglement of equal strength
(equal $V_{\rm cv}^{+}V_{\rm cv}^{-}$) can in theory achieve it.

\subsubsection{An example: Teleportation utilising biased
entanglement}

\begin{figure}[ht]
  \begin{center}
  \includegraphics[width=7.5cm]{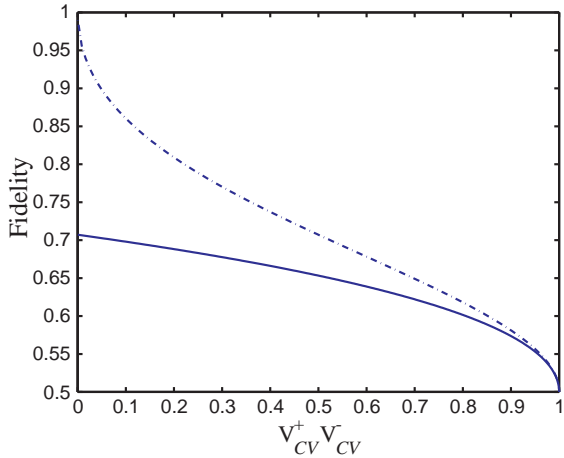}
  \end{center}
   \caption{Fidelity of teleportation of a coherent signal; both
   before (solid line) and after (dot-dashed line) redistribution of
   the entanglements quantum correlations with OPAs.}
  \label{fidelityplot}
\end{figure}

Consider entanglement generated from a single squeezed beam, for ideal 
squeezing $V_{\rm cv}^{+}V_{\rm cv}^{-}$ approaches zero and thus the 
entanglement becomes perfect.  Without specific OPA operations 
however, the maximum achievable fidelity (given by equation 
(\ref{fidelityOPA}) with $G=1$,$V^{+}_{1}=1$, and $V^{-}_{2}=0$) is 
$1/\sqrt{2}$ \cite{Loock}.  After local OPA operations with G as in 
equation (\ref{gainchoice}) the maximum fidelity (given by equation 
(\ref{Fmax})) is equal to 1.  Figure \ref{fidelityplot} shows the 
fidelity achieved first by simply using the biased entanglement and 
then by beforehand utilising our protocol to redistribute the quantum 
correlations.
For reasonably strong squeezing substantial improvements in fidelity
can be achieved via local OPA operations and indeed with just a single
squeezed beam the teleportation can be perfect.  In Ref.\cite{Loock}
the coherent state teleportation fidelity was taken as a measure of
the entanglement.  Remembering that entanglement cannot be increased
by local operations, we see that coherent state fidelity only gives a
lower bound to the entanglement strength.

\subsubsection{Another example: Teleportation of a squeezed state}
Here we consider teleportation of a squeezed state with an arbitrary 
coherent displacement.  That is the orientation and dimensions of the 
squeezed ellipse are taken as constant and known, but the coherent 
displacement is unknown.  We show that unbiased entanglement is not 
optimum in this case.

\begin{figure}[ht]
  \begin{center}
  \includegraphics[width=7.5cm]{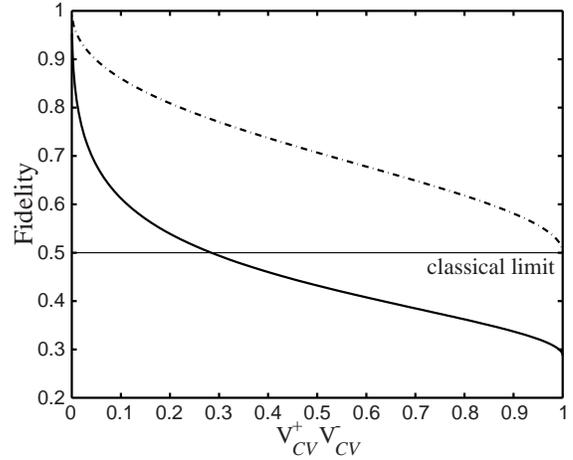}
  \end{center}
  \caption{Fidelity of teleportation of a squeezed signal ($V_{\rm
  sqz}=0.1$); utilising unbiased entanglement both with (dot-dashed
  line) and without (solid line) optimising OPA operations.}
  \label{fidelitysqueezed}
\end{figure}

The fidelity of teleportation of a minimum uncertainty amplitude
squeezed state with arbitrary coherent displacement is given by
\begin{equation}
	F = \frac{1}{\sqrt{V^{+}_{1}V^{-}_{2} + \frac{G
	V^{+}_{1}}{V_{\rm sqz}} + \frac{V^{-}_{2} V_{\rm sqz}}{G} +
	1}} \\
	\label{fidelitySQZ}
\end{equation}
where $V_{\rm sqz}$ is the variance of the squeezed quadrature of the 
signal.  Here the maximum fidelity is achieved when $G = V_{\rm 
sqz}\sqrt{V^{-}_{2}/V^{+}_{1}}$ and is equal to that given in equation 
(\ref{Fmax}); again $F=0.5$ defines the classical limit.  If unbiased 
entanglement is utilised in the teleporter $V^{+}_{1}=V^{-}_{2}$ and 
the gain simplifies to $G = V_{\rm sqz}$.  So we see that to achieve 
the maximum fidelity it is now necessary to apply parametric gain to 
the entanglement, dependent on the strength of the signal squeezing.  
A comparison between the fidelity of teleportation of a squeezed state 
with $V_{\rm sqz}=0.1$, being teleported with entanglement generated 
from two equally squeezed beams, with and without OPA operations is 
shown in figure \ref{fidelitysqueezed}.  Note that for unbiased 
entanglement as $V_{\rm cv}^{+}V_{\rm cv}^{-} \rightarrow 1$, $F 
\rightarrow (2 + V_{\rm sqz}^{-1} + V_{\rm sqz})^{-1/2}$ which is less 
than 0.5 if $V_{\rm sqz}<1$.  Thus quantum teleportation of a squeezed 
state with arbitrary coherent amplitude cannot be achieved with weak 
unbiased entanglement (see figure \ref{fidelitysqueezed} for example).

This result can be thought of as redistribution of the quantum noise
in such a way as to minimise the signal information gathered by the
amplitude and phase quadrature measurements inside the teleporter.  If
the signal is strongly amplitude squeezed then the entanglement
`noise' must be increased in the anti-squeezed quadrature to hide its
large quantum fluctuations and we can afford to decrease it in the
squeezed quadrature because the squeezed fluctuations are small.

\section{Generalisation to Multi-partite Entanglement}
We will now show that our protocol generalises to
multi-party situations.  In particular let us consider the continuous
variable GHZ state \cite {Gorb,Loock}, so-called through analogy with
the dichotomic GHZ state \cite{Green}.  We can generalise the
bipartite EPR condition to this tri-partite state by requiring that
the amplitude and phase variances of, say, beam $a$, conditioned on
measurements of beams $b$ and $c$, display an apparent violation of
the Heisenberg uncertainty relation.  The conditional variance
generalised to the three beam situation is given by
\begin{eqnarray}
V_{{\rm cv}3}^{\pm} & = & V_{a}^{\pm}-{{V_{b}^{\pm}|\langle \delta
X_{a}^{\pm} \delta X_{c}^{\pm} \rangle|^{2}+V_{c}^{\pm}|\langle \delta
X_{a}^{\pm} \delta X_{b}^{\pm} \rangle|^{2}}\over{V_{b}^{\pm}
V_{c}^{\pm}- |\langle \delta X_{b}^{\pm} \delta X_{c}^{\pm}
\rangle|^{2}}} \nonumber\\
& & -{{\langle \delta X_{a}^{\pm} \delta X_{b}^{\pm} \rangle \langle
\delta X_{a}^{\pm} \delta X_{c}^{\pm} \rangle \langle \delta
X_{b}^{\pm} \delta X_{c}^{\pm} \rangle}\over{V_{b}^{\pm} V_{c}^{\pm}-
|\langle \delta X_{b}^{\pm} \delta X_{c}^{\pm} \rangle|^{2}}}
\nonumber\\
& & -{{(\langle \delta X_{a}^{\pm} \delta X_{b}^{\pm} \rangle \langle
\delta X_{a}^{\pm} \delta X_{c}^{\pm} \rangle \langle \delta
X_{b}^{\pm} \delta X_{c}^{\pm} \rangle)^{*}}\over{V_{b}^{\pm}
V_{c}^{\pm}- |\langle \delta X_{b}^{\pm} \delta X_{c}^{\pm}
\rangle|^{2}}}
\end{eqnarray}

We can define a continuous variable GHZ violation as occurring when
$V_{{\rm cv}3}^{+} V_{{\rm cv}3}^{-}<1$.  This GHZ entanglement can be
produced by combining three squeezed states on beam splitters such that
the output fields (labelled GHZ1, GHZ2 and GHZ3) are given by
\begin{eqnarray}
\label{GHZ}
\delta  X_{{\rm GHZ}1}^{\pm} & = & \sqrt{{{1}\over{3}}} \delta
X_{1}^{\pm}-\sqrt{{{2}\over{3}}} \delta  X_{2}^{\pm}\nonumber\\
\delta X_{{\rm GHZ}2}^{\pm} & = & \sqrt{{{1}\over{3}}} \delta
X_{1}^{\pm}+\sqrt{{{1}\over{6}}} \delta X_{2}^{\pm}+
\sqrt{{{1}\over{2}}} \delta X_{3}^{\pm}\nonumber\\
\delta X_{{\rm GHZ3}}^{\pm} & = & \sqrt{{{1}\over{3}}} \delta
X_{1}^{\pm}+\sqrt{{{1}\over{6}}} \delta X_{2}^{\pm}-
\sqrt{{{1}\over{2}}} \delta X_{3}^{\pm}
\end{eqnarray}

Gorbachev and Trubilko \cite{Gorb}, and van Loock and
Braunstein\cite{Loock} consider GHZ states generated from three beams
of equal squeezing with the squeezing of beam 1 orthogonal to that of
beams 2 and 3, i.e. $V_{1}^{\pm}=V_{2}^{\mp}=V_{3}^{\mp}$.  This
indeed satisfies the condition for GHZ violation for all levels of
squeezing.  However an examination of the correlations between the
beams reveals they are biased in quadrature phase.  A
generalisation of the argument in Section IIC shows that pure GHZ
states are maximal when this noise is unbiased.  This can be arranged
here by requiring that the squeezing of input beam one is stronger
according to
\begin{equation}
V_{1}^{\pm}={{1-{V_{2,3}^{\mp}}^{2}+\sqrt{1-{V_{2,3}^{\mp}}^{2}+
{V_{2,3}^{\mp}}^{4}}} \over{V_{2,3}^{\mp}}} \label{GHZmaxcon}
\end{equation}
This then produces a maximal GHZ state in the same sense as we defined
a maximal EPR state.  That is, for a given strength of GHZ
entanglement this state has the least number of photons.

Now consider the situation of a single squeezed beam divided equally
in three.  This situation is described by equation (\ref{GHZ}) with
beam 1 squeezed, say $V_{1}^{+} <1< V_{1}^{-}$, but beam 2 and 3 in
vacuum states $V_{2}^{\pm}= V_{3}^{\pm} =1$.  The GHZ condition is
still satisfied for this state for any finite level of squeezing
provided the squeezed input beam was in a minimum uncertainty state.
This is in accordance with the conclusions of Ref.  \cite{Loock},
though the state is clearly not maximal.  However, as for the EPR
case, it is possible to create a maximal GHZ state by applying OPAs,
locally, to the three GHZ beams.  In this case the required gain of
the OPAs is
\begin{equation}
G=\frac{1}{\sqrt{3}}
\left ( \frac{V_{1}^{-}+2}{V_{1}^{+}+2} \right )^{1/2}
\end{equation}
resulting in a maximal GHZ state indistinguishable from that produced
from three squeezed states, as described by equation (\ref{GHZmaxcon}).
These techniques
can similarly be generalised to four or more parties.

\section{Conclusion}
\label{concl}

We have studied pure state EPR entanglement in which the quantum
correlations can be biased as a function of quadrature phase.  We
defined maximal EPR states as states having the most EPR entanglement for
a given number of photons or equivalently as states with minimum
photon number for a given value of $V_{\rm cv}^{+}V_{\rm cv}^{-}$.  In
particular we identified standard, unbiased EPR states produced from
pairs of squeezed beams as maximal.

Entanglement produced from a single squeezed beam with squeezed
variance $V^{+}$ was found to have the same degree of EPR correlation
as entanglement from two squeezed beams with squeezed variances
$V_{1}^{+}=V_{2}^{-}=\sqrt{V^{+}}$.  Biased entanglement is
non-maximal, however we identified local unitary operations which
could convert it to a maximal EPR state of lower photon number but the
same degree of correlation.  We examined some consequences of these
results as they apply to teleportation.

Finally we generalised our results to tri-partite entanglement.  We
identified the maximal continuous variable GHZ states.  These are not
produced from the superposition of three equally squeezed states.  We
showed that our techniques could be used to produce maximal GHZ states
from the non-maximal one produced by equally dividing a single
squeezed beam.


\begin{thebibliography}{12}

\bibitem{Bennett4}C.~H.~Bennett and D.~P.~DiVincenzo, Nature {\bf 404}, 247
(2000) and references therein.

\bibitem{Discrete} For example see P.~G.~Kwiat, K.~Mattle,
H.~Weinfurter, A.~Zeilinger, A.~V.~Sergienko and Y.~Shih, \prl {\bf
75}, 4337 (1995); J.~Brendel, N.~Gisin, W.~Tittel and H.~Zbinden, \prl
{\bf 82}, 2594 (1999); J.~D.~Franson, \pra {\bf 45}, 3126 (1992).

\bibitem{Spin} For example see A.~Peres, \pra {\bf 54}, 2685 (1996);
N.~Gisin, Phys.~Lett.~A {\bf 210}, 151 (1996).

\bibitem{Einstein} A. Einstein, B. Podolsky, N. Rosen, Phys.~Rev.
{\bf 47}, 777 (1935).

\bibitem{Reid}M.~D.~Reid and P.~D.~Drummond, \prl {\bf 60}, 2731
(1988).

\bibitem{Bra98}S.~L.~Braunstein and H.~J.~Kimble, \prl {\bf 80}, 869
(1998).

\bibitem{Lloyd} S.~Lloyd and S.~L.~Braunstein, \prl {\bf 82}, 1784
(1999).

\bibitem{Polk} R.~E.~S.~Polkinghorne and T.~C.~Ralph, \prl {\bf 83},
2095 (1999).

\bibitem{ral99} T.~C.~Ralph, \pra {\bf 61}, 010303(R) (1999).

\bibitem{Duan} L-M.~Duan, G.~Giedke, J.~I.~Cirac and P.~Zoller, \prl
{\bf 84}, 4002 (2000); L-M.~Duan, G.~Giedke, J.~I.~Cirac and P.~Zoller, \pra
{\bf 62}, 032304 (2000).

\bibitem{Bana}K.~Banaszek and K.~W\'{o}dkiewicz, \prl {\bf 82}, 2009
(1999).

\bibitem{Ral00} T.~C.~Ralph, W.~J.~Munro and R.~E.~S.~Polkinghorne,
\prl {\bf 85}, 2035 (2000).

\bibitem{Duan2} L-M.~Duan, G.~Giedke, J.~I.~Cirac and P.~Zoller, \prl
{\bf 84}, 2722 (2000).

\bibitem{Kim} Z.~Y.~Ou, S.~F.~Pereira, H.~J.~Kimble, and K.~C.~Peng,
\prl {\bf 68}, 3663 (1992).

\bibitem{Loock} P. van Loock, Samuel L. Braunstein, \prl {\bf 84},
3482 (2000).

 \bibitem{Bennett2}C.~H.~Bennett, H.~J.~Bernstein, S.~Popescu and
 B.~Schumacher, \pra {\bf 53}, 2046 (1996).

 \bibitem{Bennett3}C.~H.~Bennett, G.~Brassard, S.~Popescu,
 B.~Schumacher, J.~A.~Smolin and W.~K.~Wootters, \prl {\bf 76}, 722
 (1996).

 \bibitem{Parker} S.~Parker, S.~Bose, M.~B.~Plenio, \pra {\bf 61},
 032305 (2000).

%
%

\bibitem{Walls} D.~F.~Walls and G.~J.~Milburn, \emph{``Quantum
Optics''}, Springer-Verlag (1995).

\bibitem{SHANGXI} Yun Zhang, Hai Wang, Xiaoying Li, Jietai Jing,
Changde Xie, and Kunchi Peng, \pra {\bf 62}, 023813 (2000).

\bibitem{Silber} Ch.~Silberhorn, P.~K.~Lam, O.~Weiss, F.~Koenig,
N.~Korolkova, G.~Leuchs, quant-ph/0103002.

\bibitem{holl} M.~J.~Holland, M.~J.~Collett, D.~F.~Walls and
M.~D.~Levenson, \pra {\bf 42}, 2995 (1990).

\bibitem{Cirac} G.~Giedke, B.~Kraus, M.~Lewenstein, J.~I. Cirac, Phys.
Rev.  Lett.  87, 167904 (2001).

\bibitem{grang} Frederic Grosshans and Philippe Grangier,
\pra {\bf 64}, 010301 (2001).

\bibitem{yeo93} G.~Yeoman and S.~M.~Barnett, Journal~Mod.~Opt.  {\bf
40}, 1497 (1993).

\bibitem{Ralph}T.~C.~Ralph and P.~K.~Lam, \prl {\bf 81}, 5668 (1998).

\bibitem{Furu}A.~Furusawa, J.~L.~S{\o}rensen, S.~L.~Braunstein,
C.~A.~Fuchs, H.~J.~Kimble and E.~S.~Polzik, Science {\bf 282}, 706
(1998).
%
%
%
%
%

\bibitem{thew} R.~T.~Thew and W.~J.~Munro, \pra {\bf 63}, 030302(R)
(2001).

\bibitem{Plenio} M.~.B.~Plenio and V.~Vedral,
Contemp.~Phys  {\bf 39}, 6 (1998) and references within.

\bibitem{Bouwm}D.~Bouwmeester, J.~W.~Pan, K.~Mattle, M.~Eibl,
H.~Weinfurter and A.~Zeilinger, Nature {\bf 390}, 575 (1997).

\bibitem{Boschi}D.~Boschi, S.~Branca, F.~De Martini, L.~Hardy and
S.~Popescu, \prl {\bf 80}, 1121 (1998).

\bibitem{Fuchs}C.~A.~Fuchs and A.~Peres, \pra {\bf 53}, 2038 (1996).

\bibitem{Schu} B.~Schumacher, \pra {\bf 51}, 2738 (1995).

\bibitem{Gorb} V.~N.~Gorbachev and A.~I.~Trubilko, quant-ph/9912061
(1999).

\bibitem{Green} D.~M.~Greenberger, M.~Horne, A.~Shimony and
A.~Zeilinger, Am.~J.~Phys {\bf 58}, 1131 (1990).

\end{thebibliography}
\end{document}